# Metaverse Innovation Canvas: A Tool for Extended Reality Product/Service Development

Amir Reza Asadi[1[0000−0001−5440−8456]], Mohamad Saraee[2[0000−0002−3283−1912]], and Azadeh Mohammadi[2[0000−0003−3117−5279]]

[1] School of Information Technology, University of Cincinnati, Cincinnati, OH, 45221, USA
asadiaa@mail.uc.edu
[2] School of Science, Engineering and Environment, University of Salford, Manchester M5 4WT, UK
{m.saraee,a.mohammadi1}@salford.ac.uk

Abstract. This study investigated the factors contributing to the fail ure of augmented reality (AR) and virtual reality (VR) startups in the emerging metaverse landscape. Through an in-depth analysis of 29 failed AR/VR startups from 2016 to 2022, key pitfalls were identified, such as a lack of scalability, poor usability, unclear value propositions, and the failure to address specific user problems. Grounded in these findings, we developed the Metaverse Innovation Canvas (MIC) a tailored business ideation framework for XR products and services. The canvas guides founders to define user problems, articulate unique XR value propo sitions, evaluate usability factors such as the motion-based interaction load, consider social/virtual economy opportunities, and plan for long term scalability. Unlike generalized models, specialized blocks prompt the consideration of critical XR factors from the outset. The canvas was evaluated through expert testing with startup consultants on five failed venture cases. The results highlighted the tool's effectiveness in surfacing overlooked usability issues and technology constraints upfront, enhancing the viability of future metaverse startups.

## 1 Introduction

The concept of the metaverse has been around for decades, but it has been the subject of renewed interest and excitement in recent years as technology has advanced and virtual reality has become more accessible. The metaverse can be understood as an integrated network of spatial realities and virtual worlds, where users can engage with each other and digital elements within a three-dimensional environment, often utilizing personalized avatars [15, 5].

The emergence of AR/VR applications, services, and tools has had a pro found im-pact on shaping the metaverse into a dynamic and immersive digital realm. AR applications overlay digital information onto the real world, enrich ing our physical environment with virtual elements, while VR applications create entirely simulated environments.

Over the past few decades, there has been a significant inflow of billions of dollars in investments into the AR/VR startup ecosystem [12, 14]. However, de spite the substantial financial backing, a considerable number of these startups have faced insurmountable challenges and have ultimately failed to thrive. This trend piqued our curiosity and motivated us to embark on an in-depth explo ration of the underlying reasons behind the failure of these companies. While it is commonly acknowledged within the industry [1, 17] that technological and hardware limitations have played a pivotal role in these setbacks, we recognized the importance of delving deeper into the diverse array of factors that may have contributed to these failures. Thus, our research aimed to unravel the complex ities surrounding the demise of AR/VR startups, going beyond the prevailing notion of technological constraints.

Through an exhaustive investigation, we sought to broaden the understand ing of the dynamics at play within the AR/VR startup landscape. Our analy sis encompassed a comprehensive examination of market dynamics, user adop tion patterns, funding challenges, regulatory hurdles, and competition dynamics. Through a meticulous examination of these multifaceted factors, we aimed to attain nuanced insights into the intricate challenges confronted by these startups.

By leveraging the insights obtained from our research, we moved beyond a theoretical understanding to develop a practical solution. This solution aimed to assist founders and investors in refining their business ideas within the distinctive context of extended reality (XR) environments. While the Lean Canvas serves as a useful starting point for business model development, our analysis identi fied critical gaps when applying it to augmented and virtual reality ventures, including an inadequate emphasis on validating the problem–solution fit and ac counting for the user MIC, an innovative tool designed to facilitate the iterative refinement of business concepts. The MIC serves as a structured framework for founders and investors, enabling them to identify and address the specific needs and intricacies of XR environments. By leveraging this canvas, entrepreneurs can systematically explore various dimensions such as the user experience, the technological feasibility, the market viability, monetization strategies, and the scalability potential within the metaverse ecosystem.

Our approach of examining failed startups specifically aimed to avoid survivor ship bias, a common pitfall in entrepreneurship research where only successful cases are studied, as highlighted in [28]. Previous research has highlighted po tential pitfalls in AR/VR adoption, such as visual fatigue, cybersickness, mus cular fatigue, acute stress, and mental overload [27]. However, these studies have primarily focused on examining ergonomic and physiological risks rather than providing actionable strategies tailored to the practical context of XR startups and product development. Similarly, while the Extended by Design Toolkit [8] acknowledges challenges such as physical fatigue and increased cognitive load, it lacks specific constructs prompting developers to justify motion-based user experience choices versus less-demanding interaction methods. Specialized busi ness model frameworks such as the Blitz Canvas [25] have been proposed for specific domains like software startups, but there remains a need for tailored tools addressing the unique considerations of extended reality ventures.

Successful implementations of AR/VR have the potential to revolutionize user experiences through immersive gaming, intuitive 3D designs, interactive re mote collaboration, digitally-enhanced shopping environments, and many other applications. How-ever, achieving sustainable, scalable ventures in this space re quires addressing key issues such as hardware limitations, creating compelling XR-exclusive value propositions, justifying physical/cognitive interaction loads, and planning for interoperability and ecosystem forces [17].

This has motivated the need for tailored business modeling and design tools that can guide entrepreneurs in ideating viable, usable, and scalable AR/VR products and services by embedding these critical considerations starting in the initial stages.

This research proposed the MIC, a novel framework for developing XR prod ucts and services. The MIC addresses the limitations of existing tools such as the Lean Canvas in capturing unique considerations for XR ventures. The key ad vantages of the MIC include the following: 1) emphasizing problem solving, with blocks dedicated to framing the problem and articulating XR-exclusive value propositions; 2) incorporating usability metrics and scenario writing to com municate user experiences; 3) considering scalability issues specific to XR, such as hardware constraints; and 4) guiding ideation about the social interactions and virtual economies afforded by XR. Overall, the MIC provides a holistic, human-centered approach to ideation, enhancing the viability of XR startups by bridging the gaps between design and business modeling. Its tailored blocks prompt entrepreneurs to leverage the distinctive capabilities of immersive tech nologies. This research made several notable contributions to the understudied domain of augmented reality (AR) and virtual reality (VR) entrepreneurship. First, it identified prevalent failure factors based on an in-depth investigation of unsuccessful AR/VR startups, elucidating common pitfalls.

Second, grounded in design research and industry expertise, it presented the novel MIC as a strategic ideation toolkit tailored to addressing the unique consid erations of extended reality ventures. Third, the proposed framework prompted a human-centered, problem-driven approach to conceiving prod-ucts/services that leverage the differentiating capabilities of immersive technologies. Fourth, the methodology synthesized mixed qualitative techniques and an external eval uation to produce actionable, valid insights. Overall, this work combined an insightful analysis of real-world challenges with an innovative design output, enhancing the viability of future AR/VR startups through a tailored, reflec tive ideation process. It provided a model for human–computer interactions and design researchers seeking to bridge conceptual contributions with practical en trepreneurial impact.

## 2 Methodology

This research consisted of three main stages. In the first stage, data from domes tic and international AR/VR startups were collected and analyzed to identify the key factors contributing to their failure. Four focus group sessions were con ducted to discuss the data, and the factors were coded based on the analysis.

In the second stage, the authors applied autoethnography[21] and autobio graphical design[18] to create an innovation tool for startup founders. By lever aging their more than a decade of experience in entrepreneurship, the authors developed the MIC. This canvas serves as a framework for investors and en trepreneurs in the metaverse field. Finally, in the third stage, we recruited ex perts to evaluate five failed startups with the MIC and interviewed them about the MIC's potential for use in the AR/VR industry.

### 2.1 Stage 1: Analyzing the AR/VR Startups

The first stage of the research involved analyzing AR/VR startups. The first author's startup studio actively monitored the AR/VR industry and collected data on startups operating in various sectors, including content creation, gaming, social media, education, healthcare, tourism, and retail. The data spanned from 2016 to 2022, with 2016 being a significant milestone in AR/VR history due to the commercial success of Pokémon Go, which validated the XR market and led to the widespread adoption of computer-mediated reality technology [4].

The collected data included the Lean Canvas of the startups, which was developed based on the principles outlined in [20]. Additionally, the data encom passed the logs of usability testing sessions, led by the first author. To filter out failed startups, the authors examined the recent status of these ventures and identified 29 startups as failures. Participants were recruited through searching on LinkedIn. In order to reduce the bias, the participants were not recruited from the connections of the authors.

To evaluate the reasons behind the failure of these startups, the authors conducted focus group sessions. There were five group sessions in total, and four participants were recruited through searching on LinkedIn. In order to reduce the bias, the partici-pants were not recruited from the first connections of the authors. These focus group sessions were conducted with the same participants. The sessions took between 60 and 90 minutes and were conducted in Q1 2023.

The analysis was based on the collected data, the current status of the startups obtained from the Crunchbase directory, and the application of stress testing desirability [13]. This qualitative test focused on evaluating the problem rather than the solution, providing valuable insights into the marketability of the product or service. Finally, through a content analysis and codification, the key themes of AR/VR startup failures were identified.

The selection of AR/VR startups for the analysis was conducted through multiple channels. First, we leveraged the authors' extensive network within the startup ecosystem, reaching out to companies they had previously advised or collaborated with.

Additionally, we utilized online startup directories such as Crunchbase, An gelList, and local startup groups to identify relevant AR/VR ventures operating within the specified timeframe of 2016-2022. We then contacted the founders or leadership teams of these startups, explaining the purpose of our research and requesting their participation by sharing relevant business documentation and data. For startups that had ceased operations, we attempted to contact former employees or stakeholders to obtain access to archival materials. Throughout this process, we ensured the anonymity and confidentiality of the participating individuals and companies.

## 2.2 Stage 2: Autoethnography and Autobiographical Design

The objective of this stage was to develop a startup business-modeling framework that aimed to mitigate the mistakes made by entrepreneurs when using the Business Model Canvas (BMC) and the Lean Canvas, which have been identified as contributing factors to the failure of AR/VR startups. By recognizing and addressing the key failure factors identified in Section 3, the MIC framework sought to provide entrepreneurs with a comprehensive and effective tool for strategically planning their AR/VR startups, thus increasing their chances of success in the competitive extended reality industry.Autoethnography has gained popularity in the field of human–computer interactions in recent years [11, 19, 21].

This method is also employed in universities for conducting design research, as it allows for the collection of "creative and innovative processes" experienced by designers [16]. Autobiographical design, a form of design research, utilizes autoethnography primarily for the creation of solutions that are intended to be used extensively by their creators [18].

The primary source of data for this research included the authors' personal and extensive experience with the BMC[20] and the Lean Canvas[22]. We pro vided consulting and design services to startups and utilized strategic design toolkits such as BMC and its adaptations. Additionally, as the strategic design ers and heads of a startup studio, we shared firsthand accounts and insights gained from extensive experience working with AR/VR startups. We conducted autobiographical design interviews to gather rich and personal narratives, focus ing on the challenges, successes, and lessons learned in the context of AR/VR startup ventures.

In addition, the involvement of other authors in this research played a crucial role in facilitating ethnography and enabling data triangulation. The authors collaborated in conducting a comprehensive literature review and netnography of the startup community on Twitter, which served to triangulate the findings. This cross-validation approach enhanced the reliability and accuracy of identifying the factors contributing to failure in this study.

## 2.3 Stage 3: Evaluation

In the third stage, the MIC was evaluated through expert testing with three startup consultants. These participants were recruited based on their expertise in advising early-stage ventures and were identified by searching the websites of startup accelerators in Iran. Each consultant was presented with the Lean Canvas and background information for the five failed AR/VR startups identified in stage 1. They were then asked to fill out the MIC for each startup.

These consultants were asked to evaluate the usability and feasibility aspects of the MIC by reviewing the startup cases and providing feedback. Their insights from an expert perspective contributed to assessing the potential and usability of the proposed framework. Semi-structured interviews were conducted to gather feedback on the usefulness and usability of the canvas. The sessions were audio recorded and transcribed.

It is important to note the limitations of this approach. We did not have access to proprietary internal information for Startups 2, 3, and 4, since we did not interview their founders directly. Our perspectives on these startups were based on publicly available information, which may have lacked key insights. There was also an element of subjectivity in how we interpreted and synthesized the data on the startups based on our own prior knowledge and experiences.

Furthermore, the participants in this study had access to limited information about the startups and reviewed this information in a compressed timeframe. During the MIC testing session, three startup consultants were brought in to analyze and com-plete the canvas for five different failed AR/VR startups.

## 2.4 Analysis Method

The analysis of the collected data in this study was guided by Glaserian grounded theory methodology, which is a systematic approach for generating a theory from empirical data [7, 6]. This methodology recommends the constant comparative method, where data collection and analysis occur simultaneously, and the re searcher continuously compares incidents, codes, and categories to construct a coherent theoretical framework grounded in the data[7].

### 2.5 Ethical Considerations and Limitations

Ethical considerations were carefully addressed throughout the research pro cess. Con-fidentiality and anonymity were maintained when referencing specific startup cases or individuals within the startup studio. Informed consent was obtained to protect individuals' identifiable information and prevent potential harm to their reputation during the research. After the analysis, the collected data were destroyed. The authors main-tained reflexivity throughout the study by acknowledging their positional ties to the startup community and the poten tial influence it may have had on their observations and interpretations.

## 3 Key Failure Factors of AR/VR Startups

During the analysis of AR/VR startups, several key failure factors were identi fied, shedding light on the challenges that led to their downfall. These factors contributed to a deeper understanding of the reasons behind the failures of these startups. The following factors emerged as significant contributors to the failure of AR/VR startups:

### 3.1 Short Timeframe and Lack of Future Planning

Many AR/VR startups faced difficulties due to a short-sighted approach and a lack of clear plans for future trends. The rapidly evolving nature of the AR/VR industry demands foresight and adaptability. Startups that failed often lacked the ability to anticipate market shifts, technological advancements, and changing consumer preferences, ultimately leading to their downfall.

This factor was observed in 18 out of the 29 failed startups analyzed. Many of these companies focused on short-term trends or gimmicks without accounting for longer-term market shifts or technological advancements.

### 3.2 Lack of Scalability and Hardware Limitations

The lack of scalability in the AR/VR industry can be attributed to the limited availability of AR/VR head-mounted displays (HMDs), preventing the creation of a mass market service. The shortage of AR/VR HMDs hampers the scal ability of startups, as there is not yet a sufficient consumer base to support widespread adoption. The hard-ware limitations and the industry's challenge in providing affordable, accessible, and high-quality devices have further hindered the scalability of AR/VR startups.

A total of 22 of the 29 startups struggled with scalability issues stemming from the limited availability and high cost of AR/VR headsets and other en abling hardware during their active years. Their solutions could not achieve mass consumer adoption due to these hardware constraints.

Quantifying this constraint, only 8.1 million AR/VR headsets were shipped globally in 2018 (source: IDC Worldwide Quarterly AR/VR Device Tracker). With limited device penetration and high costs, such as USD 599 for the Oculus Rift, 15 of the 29 failed startups (52%) directly cited hardware availability and pricing as a key factor hampering the scalability of their AR/VR solutions.

### 3.3 Lack of Usability and Value Proposition

Usability plays a crucial role in the success of AR/VR startups. The startups that neglected the importance of intuitive and user-friendly experiences faced significant challenges. Additionally, the lack of a compelling value proposition, considering non-AR/VR alternatives, contributed to the failure of some startups. When users could find similar solutions outside the AR/VR environment that were more accessible, more cost-effective, or easier to use, the value proposition of the AR/VR startups became diminished.

While [27] comprehensively reviewed the risks of visual fatigue, cybersick ness, and other physical discomforts that undermine the usability and value of immersive technologies, their work stopped short of outlining actionable strate gies tailored to the practical context of XR startups and product development. Specifically, their review did not translate their insights on ergonomic pitfalls into a coherent framework for mitigating issues starting from the early stages of busi ness modeling and design. Our research built upon their analysis by proposing the MIC as an entrepreneurial toolkit for embedding a consideration of critical usability and user experience (UX) factors directly into the ideation process. While [23] systematic review highlighted the usability and physical burden is sues with augmented reality in health sciences education, their work did not provide an actionable framework for addressing these concerns during business ideation for AR/VR ventures. Our proposed MIC aims to bridge this gap by directly incorporating human-centered experience scenarios and prompting the justification of motion-based interactions from the start.

A total of 17 of the failed startups exhibited poor usability of their AR/VR offerings, with unintuitive or cumbersome user experiences that diminished the value proposition. Additionally, 16 of them failed to clearly articulate a com pelling value proposition that justified the use of AR/VR over non-immersive alternatives.

An AR home decor app received widespread complaints about poor usability, with a 2.1/5 rating on the App Store. The users cited issues such as "keeps crashing" and "difficult to place objects naturally." Without a compelling value add over browsing decor websites, the app struggled with engagement.

### 3.4 Motion-Based Interactions and Physical Load

Some failed AR/VR startups struggled with motion-based interactions and phys ical load. While these startups offered innovative motion-based experiences, the physical demands and discomfort associated with prolonged use did not justify the value propositions for users. The disparity between the expected advantages and the physical burden on users eroded the feasibility and attractiveness of their offerings.

Although the Extended by Design Toolkit [8] acknowledges physical and per ceptual strain as challenges in XR, it lacks constructs prompting developers to justify motion-based UX choices versus less-demanding interaction methods. Our canvas guides teams to weigh the benefits and burdens of natural user interfaces while considering the long-term feasibility. Although the review by Buchner et al. [3] described the physical burdens associated with augmented reality, their analysis lacked guidelines for mitigating these demands during design and de velopment. Studies have shown that demanding augmented reality experiences increase the strain, reducing the com-fort and performance over time [2]. Simi larly, research has found significant fatigue from the extended use of augmented reality glasses compared to mobile devices, limiting their feasibility for daily tasks. Similar to the usability issues, the problems with motion-based interac tions and physical load for AR/VR systems [9] are not being accounted for in current generalized startup planning models such as the Lean Canvas, whereas our proposed MIC provides specific fields to prompt the consideration of these critical factors.

The experiences of 14 of the startups imposed high physical and cognitive loads on their users through excessive motion-based interactions. The benefits did not out-weigh the burdens of these demanding natural user interface require ments.

An AR productivity startup failed after the users reported physical strain from the extended motion-tracking required for 3D modelling. An App Store review stated: "It’s innovative but using hand gestures for more than 10 minutes is very tiring. I can’t see people using this daily." The startup could not justify the interaction load versus mouse/keyboard inputs.

### 3.5 Failure to Address a Clear Problem

Successful startups often solve tangible problems or fulfill specific needs. How ever, some AR/VR startups failed to address a clear problem or failed to ef fectively communicate the problem they were solving. This lack of focus on a defined problem reduced the perceived value of their offerings and hindered their ability to attract and retain customers.

A total of 11 of the analyzed startups did not have a well-defined central problem that their AR/VR solution effectively solved for users. A lack of focus on addressing real user needs undermined their perceived value.

A VR startup for remote communication had no clear user workflow or pain point that their solution addressed better than video calls. As one reviewer stated: "It’s novel tech but I don’t understand when I would actually use this instead of a Zoom call."

By recognizing these key failure factors, future AR/VR startups can learn from the mistakes of their predecessors and take proactive measures to mit igate these challenges. Addressing issues related to long-term planning, hard ware limitations, usability, the value proposition, motion-based interactions, and problem-solving can significantly increase the chances of success in the compet itive AR/VR landscape.

## 4 Development of the Metaverse Innovation Canvas

The Lean Canvas, as a perfect one-page business plan template for brainstorming business ideas, was introduced in the Running Lean book [13] based on an ex tension of the Business Model Canvas [20], and it has helped a countless number of startups around the globe. However, since the Lean Canvas[22] is a general purpose framework, entrepreneurs may not use it correctly. The following is a list of common myths, mistakes, and difficulties that AR/VR startup founders usually encounter when trying to deconstruct their business assumptions with the Lean Canvas.

### 4.1 Myths, Difficulties & Mistakes

A common mistake during idea development is only considering digital and XR alternatives, while neglecting non-digital options. For example, when conceptu alizing an XR shopping app, the founders may focus solely on existing digital solutions such as online shopping websites. However, they should also consider brick-and-mortar stores as a non-XR alternative that users may currently uti lize. The tendency is to view the landscape strictly within the XR application space, but founders need to look beyond this and analyze all potential digital and non-digital alternative solutions to the problem they are trying to solve.

Bringing an application to XR is a value proposition. This is wrong; it might pro-vide an early blessing in an untapped market, but XR brings extra costs and additional physical and even cognitive loads [29] for users, so these extra costs need to be justified by providing value propositions for end users. Pokémon Go motivated [26] its users to move outside and follow location-based interaction practices because the users were collecting Pokémon cards before that. Niantic did not need to justify this extra load for the users, since years of marketing by Nintendo had already accomplished this for them.

Problem definition is the toughest job for every entrepreneur and product designer, and the problem is further complicated by the fact that most peo ple do not have clear understanding of what can be considered a problem. For example, the case of Angry Birds may not immediately seem relevant to the prob lem–solution definition. How-ever, the main issue that Angry Birds addresses is people's need to maintain their attention on something, and this game accom plishes this in an entertaining and user-friendly manner.

The primary aim of the Lean Canvas is to help founders validate their idea as soon as possible, and for AR/VR startup founders, it is ideal to refine their ideas into viable, desirable, and feasible solutions. However, many AR/VR solu tions struggle to achieve feasibility or scalability in a short time due to the lack of advancements in enabling technologies. Therefore, the perfect business plan for AR/VR startups should consider the impacts of predictable future market events.

The entreprenurs have demonstrated two common mistakes in desiging star tups: Large customer segment, and lack of attention to usability considerations: Entrepreneurs tend to expand their customer segments and write more items in this segment; it may sound more interesting for some investors because it can increase the size of their market and the valuation of the company, but it can be a fatal mistake in the XR industry. This is because it increases the development costs exponentially and will postpone idea validation for a long period. It would be wise to limit the early adopters and find a niche market for the minimum viable products. Thiel and Masters[28] argued that only monopoly profits can help a business to survive, since a monopoly provides the profits to continue the innovation.

Postponing the usability considerations until the minimum marketable prod uct is reached. UX issues in XR environments are far more complex than screen based applications. Handheld AR can cause hand fatigue [3], and users cannot use face com-puters such as AR/VR HMDs for multiple hours. These consider ations directly im-pact the sustainability of the business model. In response to these concerns, we developed a one-page business plan tem plate specifically tailored for XR startups, known as the Metaverse Innovation Canvas, as depicted in Figure [?]. The MIC comprises five distinct types of blocks, with each serving a unique purpose: problem (red blocks), solution (green blocks), usability (blue blocks), viability (yellow blocks), and future scalability (purple blocks).

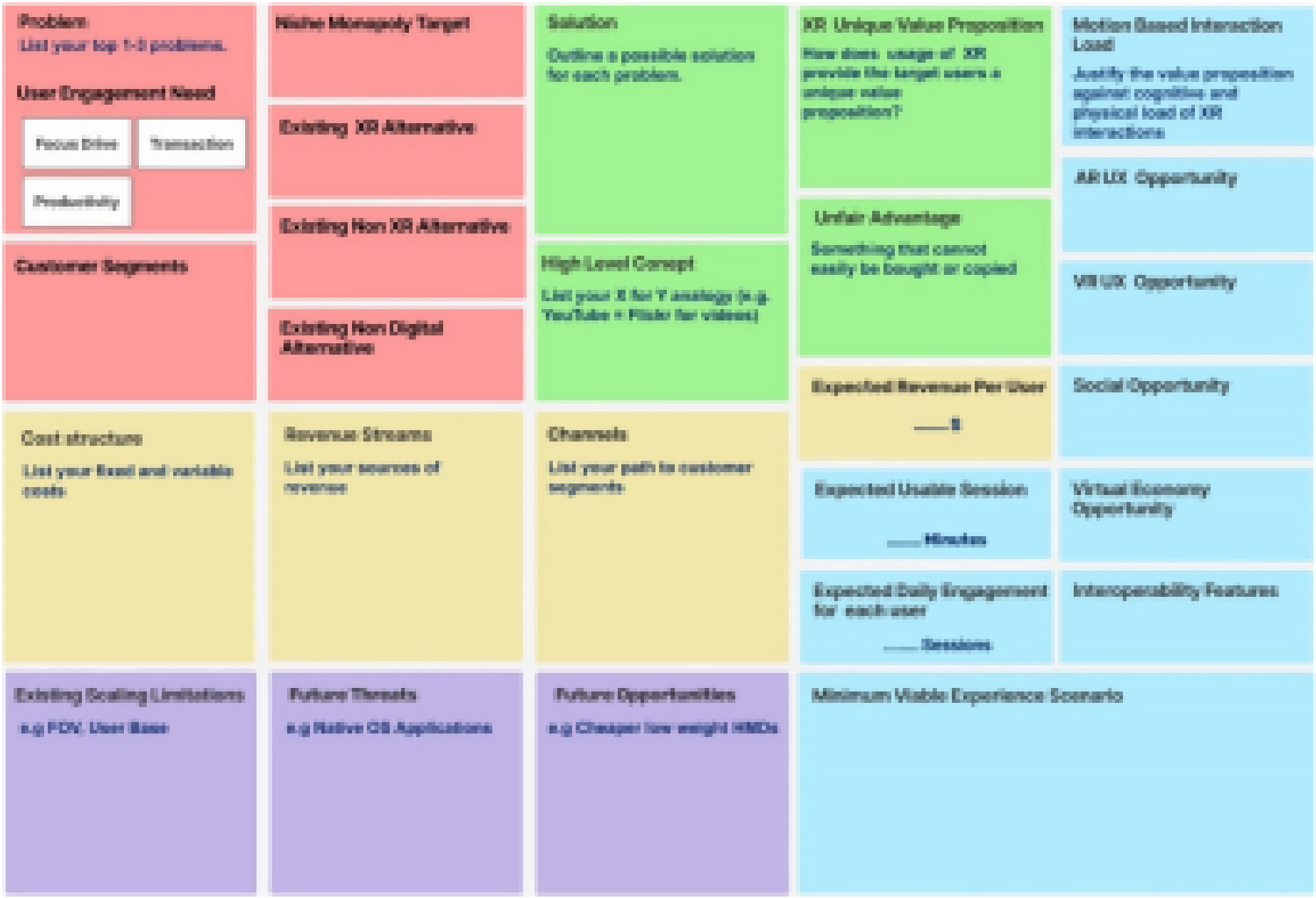

Fig. 1. Metaverse Innovation Canvas

## 4.2 Problem

The problem section includes blocks that describe a problem and the people who have the problem. In order to help designers and founders define their problems more effec-tively, we added user engagement goal checkboxes to the problem section, which included focus drive, productivity, and transaction. We adopted this from an article about the Three Games of Customer Engagement Strategy [10].

Focus drive represents users' inherent need to direct their attention and con centration towards specific subjects or activities. It reflects individuals' natural motivation or inclination to concentrate their mental resources on particular tasks or areas of interest. Solutions in this category address this need by offering value in the form of information, entertainment, or self-expression. Netflix, so cial media, and video games are examples of solutions that address focus drive problems [10]. The myths around only considering digital/XR alternatives and the thought that bringing an app to XR is a value proposition are addressed by separating "existing XR alternatives" from "existing non-XR alternatives" and "existing non-digital alter-natives." This prompts founders to consider a broader set of solutions beyond just XR. The difficulty in defining the core problem is mitigated by the "user engagement goal" checkboxes that guide problem framing in terms of the focus drive, productivity, or transaction needs.

### 4.3 Solution

The concept of a unique value proposition has been adapted to the context of XR, resulting in the XR unique value proposition. This shift is essential for startup designers to recognize and emphasize the value propositions that users can exclusively experience through XR solutions, distinguishing them from non XR alternatives.

In the XR landscape, it is crucial to identify and leverage the immersive and trans-formative qualities that XR technologies offer. Startups need to go beyond replicating existing non-XR solutions and instead focus on harnessing the unique capabilities of XR to provide novel and engaging experiences that cannot be achieved through traditional means.

The "XR unique value proposition" block directly counters the myth that simply bringing something to XR is valuable by pushing founders to articulate the value enabled specifically through XR capabilities.

### 4.4 Usability

We aimed to minimize the usability ideation in a text-based manner by creating the following blocks. Filling all the fields is not mandatory for each startup, but it can give a quick overview about the position of a startup in the metaverse landscape. Motion-based interaction load:

The motion-based interaction load encompasses the physical and cognitive demands imposed on users as they interact with XR experiences using gestures and body movements. The value proposition of motion-based inter-actions lies in their ability to provide a more immersive and intuitive user experience, en hancing engagement and enabling new possibilities in fields such as training simulations and gaming. However, it is crucial to strike a balance between the benefits and the associated load, considering factors such as user comfort and the learning curve. By justifying their value proposition and addressing chal lenges, startups can leverage motion-based interactions to create compelling XR experiences that captivate users and drive the adoption of this transformative technology.

– AR UX opportunity: How does this solution help users in a semi-immersive environment?

– VR UX opportunity: How does this solution help users in a fully immer sive environment?
– Social opportunity: How does this solution create value for users via social interactions?
– Virtual economy opportunity: How does using the virtual world econ omy paradigm create value for customers?

– Interoperability features: How does the solution adopt interoperability or achieve the interoperability needs of users?
– Minimum viable experience scenario: This is a written form of a user storyboard that explains an ideal usage story in a short manner, which may include all or some pieces of the usability elements.

The motion-based interaction load addresses the mistake of deferring us ability considerations by prompting the upfront justification of motion interfaces versus physical/cognitive burdens. The "AR UX opportunity" and "VR UX op portunity" guide the ideation of AR and VR use cases distinctly. The "social opportunity" and "virtual economy opportunity" promote the consideration of the novel social and economic models afforded by XR worlds. The "minimum viable experience scenario" narrative description helps convey and refine the envisioned user experience holistically.

### 4.5 Viability elements

The viability blocks of the MIC encompass essential elements such as channels, costs, and revenue streams, which were derived from the Lean Canvas framework. These blocks allow startups to strategically plan and analyze their distribution channels, estimate the costs associated with their metaverse ventures, and iden tify potential revenue streams. These elements inherited from the Lean Canvas address the difficulties around premature scaling by starting with a minimally viable offering before expanding.

### 4.6 Future Scalability

The future scalability block within the MIC includes various crucial aspects, including the existing scalability limitations, future threats, and future oppor tunities.

By considering these elements, founders can assess the potential for scaling their metaverse ventures over time. This block enables startups to identify and understand any existing limitations or challenges that may hinder their scalabil ity, allowing them to proactively strategize and overcome such obstacles. Addi tionally, by considering future threats and opportunities, startups can anticipate the market dynamics, emerging technologies, and changing user preferences that may impact the scalability of their metaverse solutions. This forward-thinking approach empowers startups to position themselves for future growth, adapt to evolving trends, and seize new opportunities within the rapidly evolving meta verse landscape.

The "existing scalability limitations" and "future threats" counter the myth and difficulty of short-term planning by pushing founders to forecast the ob stacles to long-term growth. "Future opportunities" guides the consideration of how evolving technology and trends can enable the XR solution to be scaled over time.

### 4.7 Measurable Metrics

To account for the critical role of usability in the profitability of XR products and services, we incorporated predefined metrics into the MIC. These metrics included the expected usable session (minutes), representing the anticipated du ration of each user session; the expected daily engagement for each user (ses sions), capturing the project-ed frequency of user engagement; and the expected revenue per user, providing in-sights into the anticipated revenue generated from individual users.

## 5 Testing the model with Startup Experts

This study included an analysis of the five startups presented in Table 1: two developed a shopping application, another specialized in gaming, one ventured into social networking, and the final one concentrated on productivity solutions.

The consultants had no prior knowledge of the specific reasons behind the failures of these startups. They were provided with the summary, main usage scenario, and start and stop dates of these ventures.

A common theme emerged among all the participants that emphasized the significance of XR value propositions. It was collectively observed that three out of the five failed startups did not offer value propositions that leveraged the unique capabilities of XR systems, missing out on the potential benefits that the technology could provide. This highlights the importance of creating value propositions that are exclusive to XR experiences in order to fully capitalize on the potential of the technology.

The shopping app, VR social network, AR social network, and productivity app startups faced challenges that justified the value of their solutions against the cognitive and physical demands placed on the users. All the participants emphasized the need to justify motion-based interactions. One consultant sug gested that XR startups should avoid brief, buzzword-focused pitches centered on being "AR/VR," and in-stead clearly articulate their unique value proposi tion. For example, the shopping app only implemented an AR display feature. As its competitors had integrated similar in-room display capabilities, this startup lacked a defensible advantage. One participant stated that pointing out future threats prompted the founders to think beyond the short-term trends and con sider sustainable differentiation.

One of the participants expressed appreciation for the modification made to the existing alternative block, recognizing its ability to stimulate ideation and inspire new ideas during the design process. They acknowledged the value of this modification in expanding the possibilities for innovations within the XR space. When the designers were queried about the usability blocks, there was a general consensus among the participants that these blocks are crucial for effectively communicating ideas prior to the development phase. However, they also noted that filling this section may require a solid understanding of AR/VR technol ogy, suggesting that founders may need the assistance of an XR specialist to accurately complete this field. This highlights the technical knowledge required to address usability concerns and design effective user experiences within the AR/VR domain.

Table 1. Analysis of five startups

| Categor y | | Start Year Stop Year Summar y | |
|---|---|---|---|
| Startup #1 Shop ping | Q3 2022 | Q2 2022 | This startup was an AR android application for buying home decor.The revenue model was based on selling commissions. The app included a Web back end, and shop owners were able to list their products. The shop included more than 1000+ products and achieved 100,000+ installations. |
| Startup #2 Shop ping | Q2 2019 | Q2 2021 | This startup was a VR app with 10,000+ installations that simulatedshopping malls. The app's revenue model relied on selling subscriptions to shop owners. The users needed to have Google Cardboard VR to use the app. |
| Startup #3 Social Networ k | Q2 2018 | Q3 2021 | This startup was an AR-only location-based social network that allowedpeople to meet other people around them and evaluate. It crossed 5K users. The app's revenue model relied on selling ad-spaces for sending AR gifts and AR. The app was available for phones, tablets and never models, and it was installed on more than 100,000 devices. |
| Startup #4 Gam ing | Q1 2018 | Q1 2022 | This startup developed a location-based AR mobile game available oniOS and Android, generating 200,000+ installation. Revenue was gen erated through in-app purchases. Despite a successful launch, the game faced a swift decline in user retention. The gameplay required physical movement, incorporating the player's actual locomotion as a core me chanic. |
| Startup #5 Pro ductivit y | Q3 2017 | 2021 | This app was released for iPhone and iPad, and most features requiredthe constant use of AR mode. The app aimed to facilitate the organiza tion of to-do lists by augmented reality by animating 3D objects. The users were able to use the app for free and it crossed 20,000 installa tions. |

Furthermore, two participants emphasized that the lack of scalability within a short timeframe was a prominent failure factor for two of the startups. One participant pro-posed the addition of a timeline block to the MIC to address this concern. By incorporating a timeline, startups would be prompted to consider and plan for the scalability of their ventures over time, recognizing the potential threats and opportunities that may arise in the future.

Overall, the engagement of the startup consultants in completing the MIC high-lighted key areas of focus, such as the XR value propositions, usability considerations, and scalability.

## 6 Limitations and Future work

One limitation of this study was the sole reliance on qualitative data from startup consultants to evaluate the MIC. While the feedback of these consultants pro vided valuable insights, incorporating quantitative metrics to objectively mea sure key factors such as the XR value propositions, usability considerations, and scalability could further strengthen the evaluation. Future research should explore the development and validation of quantitative instruments to assess these critical elements within the context of XR business modeling. By combin ing qualitative expert feedback with quantitative measurements, a future study could triangulate the findings and provide a more comprehensive understanding of the canvas's efficacy in enhancing the viability of metaverse startups.

Another limitation of this study was the lack of an analysis on the success factors for thriving AR/VR startups. While our investigation of failed ventures provided valuable insights into common pitfalls, incorporating an examination of successful cases could further enrich the findings and provide a more com prehensive understanding of the critical factors influencing a startup's success in the AR/VR domain. Future research could extend this work by analyzing prosperous AR/VR startups and integrating the identified success factors into the MIC framework.

The analysis of market dynamics and user adoption patterns in the AR/VR

industry may have been oversimplified in this study. A more in-depth explo ration of these factors could yield additional insights and nuances that were not captured in the cur-rent scope of this research. The data collection for this study was conducted in the previous year, and may not have fully captured the rapidly evolving nature of the AR/VR industry. As new technologies and market trends emerge, the identified failure factors and the applicability of the MIC may require further validation and refinement.

Expanding the research scope to include AR/VR startups from diverse cul tural and geographical contexts could enhance the understanding of region specific factors that influence startup success or failure. As new technologies such as blockchains, artificial intelligence, and the Internet of Things[24] continue to shape the metaverse landscape, future research could explore the integration of these advancements into the MIC, enabling a more comprehensive and future proofed ideation process. Conducting empirical studies to validate the efficacy of the MIC in real-world scenarios could provide quantitative evidence of its impact on the success rates of AR/VR startups.

## 7 Concluding Remarks

This research made several notable contributions to the understudied domain of augmented reality (AR) and virtual reality (VR) entrepreneurship. First, it identified prevalent failure factors based on an in-depth investigation of unsuc cessful AR/VR startups, elucidating common pitfalls that can lead ventures in this space to fail.

Understanding these factors can provide valuable lessons for aspiring en trepreneurs seeking to increase their chances of success. Second, grounded in design research methods such as autoethnography, autobiographical design, and insights from industry expertise, this study presented the novel MIC as a strate gic ideation toolkit tailored to addressing unique considerations of extended reality ventures.

This specialized canvas guides founders through prompts that encourage the consideration of critical factors such as the user experience, the technological feasibility, the problem–solution fit, and the business viability starting from the earliest stages. Third, the proposed MIC framework prompts a human-centered, problem-driven approach to conceive products and services that leverage the differentiating capabilities of immersive technologies such as AR and VR.

Its tailored blocks push entrepreneurs to ideate solutions that provide value propositions exclusive to XR, going beyond simply replicating non-XR expe riences. Fourth, the research methodology synthesized mixed qualitative tech niques such as focus groups, content analyses, and external evaluations by startup consultants to produce actionable and valid insights. The triangulation of per spectives enhanced the reliability and applicability of the findings.

Overall, this work combined an insightful analysis of the real-world chal lenges faced by failed AR/VR startups with an innovative design output—the MIC—aimed at enhancing the viability of future ventures in this space through a tailored, reflective ideation process. The MIC provides a model for human–computer
interactions and design researchers seeking to bridge conceptual contributions with practical entrepreneurial impact to drive the adoption of novel technolo gies.

## References

1. Alford, H.: 3 industries that will be disrupted by pokémon go and ar. Medium (2016),
2. Bambušek, D., Materna, Z., Kapinus, M., Beran, V., Smrž, P.: Handheld aug mented reality: Overcoming reachability limitations by enabling temporal switch ing to virtual reality. In: Proceedings of the 2022 17th ACM/IEEE International Conference on Human-Robot Interaction (HRI). pp. 698–702. IEEE (2022)
3. Buchner, J., Buntins, K., Kerres, M.: The impact of augmented reality on cognitive load and performance: A systematic review. Journal of Computer Assisted Learning 38(2), 285–303 (2022)
4. Chen, D.: The metaverse is here...but is the hardware ready? Spiceworks (2022), https://www.spiceworks.com/tech/hardware/guest-article/the-metaverse is-here-but-is-the-hardware-ready/
5. Dionisio, J.D.N., Burns III, W.G., Gilbert, R.: 3d virtual worlds and the metaverse: Current status and future possibilities. ACM Computing Surveys (CSUR) 45(3), 1–38 (2013)
6. Glaser, B.G.: Theoretical sensitivity. Sociology Press, Mill Valley, CA (1978) 7. Glaser, B.G., Strauss, A.L.: The discovery of grounded theory: Strategies for qual itative research. Aldine, Chicago (1967)
8. Gomes, A., Figueiredo, L., Correia, W., Teichrieb, V., Quintino, J., da Silva, F.Q., Santos, A., Pinho, H.: Extended by design: A toolkit for creation of xr experi ences. In: Proceedings of the 2020 IEEE International Symposium on Mixed and Augmented Reality Adjunct (ISMAR-Adjunct). pp. 57–62. IEEE (2020)
9. Graser, S., Nielsen, L.H., Böhm, S.: Factors influencing the user experience of mo bile augmented reality apps: An analysis of user feedback based on app store user reviews. In: Godulla, A., Böhm, S. (eds.) Digital Disruption and Media Transfor mation, pp. 109–128. Springer International Publishing (2023)
10. Hegde, S.: Defining your customer engagement strategy: Which game are you play ing? (2023), https://amplitude.com/blog/customer-engagement-strategydefining your-customer-engagement-strategy-which-game-are-you-playing
11. Heshmat, Y., Neustaedter, C., DeBrincat, B.: The autobiographical design and long term usage of an always-on video recording system for the home. In: Proceedings of the 2017 Conference on Designing Interactive Systems. pp. 675–687 (2017)
12. Matney, L.: Investments in vr/ar have already hit $1.1 billion in 2016 (2016), https://techcrunch.com/2016/03/07/investments-in-vrar-have-already-hit-1-1-

billion-in-2016/
13. Maurya, A.: Running Lean. O'Reilly Media, Inc., 3 edn. (2022) 14. Metinko, C.: Vr/ar investments increase just as metaverse talk heats up—but that may not be the only reason (2022), https://news.crunchbase.com/startups/metaverse-augmented-reality-virtual reality-investment/
15. Montes: What is spatial computing and how is it revolutionizing our world? (2022), https://www.verses.ai/blogs/what-is-spatial-computing-and-how is-it-revolutionizing-our-world
16. Munro, A.J.: Autoethnography as a research method in design research at univer sities. In: Proceedings of the 20/20 Design Vision Conference. p. 156 (2011) 17. Naess, M.: The metaverse in retail: A game-changer that's not ready yet (2022), https://venturebeat.com/datadecisionmakers/the-metaverse-in-retail a-game-changer-thats-not-ready-yet/

18. Neustaedter, C., Sengers, P.: Autobiographical design: What you can learn from designing for yourself. Interactions 19(6), 28–33 (2012)
19. O'Kane, A.A., Rogers, Y., Blandford, A.E.: Gaining empathy for non-routine mo bile device use through autoethnography. In: Proceedings of the SIGCHI Confer ence on Human Factors in Computing Systems. pp. 987–990 (2014)
20. Osterwalder, A., Pigneur, Y.: Business Model Generation: A Handbook for Vision aries, Game Changers, and Challengers, vol. 1. John Wiley & Sons, Hoboken, NJ, USA (2010)
21. Rapp, A.: Autoethnography in human-computer interaction: Theory and prac tice. In: New Directions in Third Wave Human-Computer Interaction: Volume 2-Methodologies, pp. 25–42. Springer, Cham, Switzerland (2018)
22. Reis, E.: The Lean Startup. Crown Business, New York, NY, USA (2011) 23. Rodríguez-Abad, C., Alonso-Atienza, F., Garcillán-Izquierdo, S., Reina-Muñoz, M., Lara-Del Río, R., León-Brito, N., García-González, J.M., Álvarez Salvago, F., Fernández-Domínguez, I., Salgado-Cifuentes, Y., et al.: A systematic review of augmented reality in health sciences: A guide to decision-making in higher educa tion. International Journal of Environmental Research and Public Health 18(8), 4262 (2021)
24. Sangeethaa, S., Jothimani, S.: Blockchain in the metaverse. In: Cultural Marketing and Metaverse for Consumer Engagement, pp. 51–70. IGI Global, Hershey, PA, USA (2023)
25. Shanbhag, N., Pardede, E.: The blitz canvas: A business model in novation framework for software startups. Systems 10(3), 58 (2022). https://doi.org/10.3390/systems10030058
26. Singh, M.: Pokemon go changes everything and nothing for arvr (2016), https://techcrunch.com/2016/08/12/pokemon-go-changes-everything-and nothing-for-arvr/
27. Souchet, A.D., Lourdeaux, D., Pagani, A., Rebenitsch, L.: A narrative review of immersive virtual reality's ergonomics and risks at the workplace: Cybersickness, visual fatigue, muscular fatigue, acute stress, and mental overload. Virtual Reality 27, 19–50 (2023)
28. Thiel, P., Masters, B.: Zero to One: Notes on Startups, or How to Build the Future. Currency, New York, NY, USA, 1 edn. (2014)
29. Van Der Land, S., Schouten, A.P., Feldberg, F., van Den Hooff, B., Huysman, M.: Lost in space? cognitive fit and cognitive load in 3d virtual environments.